\pdfoutput=1

\documentclass{PNAStwo}

\usepackage{graphicx}
\usepackage{rotate}
\usepackage{epsfig}
\usepackage{cite}
\renewcommand{\bf}{}
\renewcommand{\underline}{}

\usepackage{amssymb,amsfonts,amsmath}

\begin{document}



\conflictofinterest{No conflicts of interest.}



\footcomment{Author Contributions: D.H. developed the concept of
this study, supervised it, and wrote the manuscript. W.Y. performed
the computer simulations and prepared the figures.}

\title{The outbreak of cooperation among success-driven individuals under noisy conditions}

\author{Dirk Helbing\thanks{To whom correspondence should be addressed. E-mail: dhelbing@ethz.ch}\affil{1}{ETH Zurich, Chair of Sociology, in particular of Modeling and Simulation, UNO D11, Universit\"atstr. 41, 8092 Zurich, Switzerland} and Wenjian Yu\affil{1}{}}

\contributor{Submitted to Proceedings of the National Academy of
Sciences of the United States of America}

\maketitle

\begin{article}

\begin{abstract}
According to Thomas Hobbes' \textsl{Leviathan} (1651, English ed.:
Touchstone, New York, 2008), ``the life of man [is] solitary, poor,
nasty, brutish, and short'', and it would need powerful social
institutions to establish social order. In reality, however, social
cooperation can also arise spontaneously, based on local
interactions rather than centralized control. The self-organization
of cooperative behavior is particularly puzzling for social dilemmas
related to sharing natural resources or creating common goods.  Such
situations are often described by the prisoner's dilemma. Here, we
report the sudden outbreak of predominant cooperation in a noisy
world dominated by selfishness and defection, when individuals
imitate superior strategies and show success-driven migration. In
our model, individuals are unrelated, and do not inherit behavioral
traits. They defect or cooperate selfishly when the opportunity
arises, and they do not know how often they will interact or have
interacted with someone else. Moreover, our individuals have no
reputation mechanism to form friendship networks, nor do they have
the option of voluntary interaction or costly punishment. Therefore,
the outbreak of prevailing cooperation, when directed motion is
integrated in a game-theoretical model, is remarkable, particularly
when random strategy mutations and random relocations challenge the
formation and survival of cooperative clusters. Our results suggest
that mobility is significant for the evolution of social order, and
essential for its stabilization and maintenance.
\end{abstract}

\keywords{spatial games | evolution of cooperation | pattern
formation}


\dropcap{W}hile the availability of new data of human mobility has
revealed relations with social communication patterns
\cite{Barabasi} and epidemic spreading \cite{Brockmann}, its
significance for the cooperation among individuals is still widely
unknown. This is surprising, as migration is a driving force of
population dynamics as well as urban and interregional dynamics
\cite{Batty,Weidlich,Pumain}.
\par
Below, we model cooperation in a game-theoretical way
\cite{vonNeumann,Axelrod,Skyrms}, and integrate a model of stylized
relocations. This is motivated by the observation that individuals
prefer better neighborhoods, e.g. a nicer urban quarter or a better
work environment. To improve their situation, individuals are often
willing to migrate. In our model of success-driven migration,
individuals consider different alternative locations within a
certain migration range, reflecting the effort they are willing or
able to spend on identifying better neighborhoods. How favorable a
new neighborhood is expected to be is determined by test
interactions with individuals in that area (``neighborhood
testing''). The related investments are often small compared to the
potential gains or losses after relocating, i.e. exploring new
neighborhoods is treated as  ``fictitious play''.  Finally,
individuals are assumed to move to the tested neighborhood that
promises to be the best.
\par
So far, the role of migration has received relatively little
attention in game theory
\cite{Hegselmann,Epstein,NegMob1,NegMob3,Dieckmann,Mobility1,Mobility2,Akt},
probably because it has been found that mobility can undermine
cooperation by supporting defector invasion \cite{NegMob1,NegMob3}.
However, this primarily applies to cases, where individuals choose
their new location in a random (e.g. diffusive) way. In contrast,
extending spatial games by the specific mechanism of {\it
success-driven} migration can support the survival and spreading of
cooperation. As we will show, it even promotes the spontaneous {\it
outbreak} of prevalent cooperation in a world of selfish individuals
with various sources of randomness (``noise''), starting with
defectors only.

\section{Model}
\par
Our study is carried out for the prisoner's dilemma game (PD). This
has often been used to model selfish behavior of individuals in
situations, where it is risky to cooperate and tempting to defect,
but where the outcome of mutual defection is inferior to cooperation
on both sides  \cite{Axelrod,Five}. Formally, the so-called
``reward'' $R$ represents the payoff for mutual cooperation, while
the payoff for defection on both sides is the ``punishment'' $P$.
$T$ represents the ``temptation'' to unilaterally defect, which
results in the ``sucker's payoff'' $S$ for the cooperating
individual. Given the inequalities $T > R > P > S$ and $2R > T+S$,
which define the classical prisoner's dilemma, it is more profitable
to defect, no matter what strategy the other individual selects.
Therefore, rationally behaving individuals would be expected to
defect when they meet {\it once}. However, defection by everyone is
implied as well by the game-dynamical replicator equation
\cite{Epstein}, which takes into account imitation of superior
strategies, or payoff-driven birth-and-death processes. In contrast,
a coexistence of cooperators and defectors is predicted for the
snowdrift game (SD). While it is also used to study social
cooperation, its payoffs are characterized by $T>R>S>P$ (i.e. $S>P$
rather than $P>S$).
\par
As is well-known \cite{Five}, cooperation can, for example, be
supported by repeated interactions \cite{Axelrod}, by intergroup
competition with or without altruistic punishment
\cite{Multilevel1,Punish,Gintis1}, and by network reciprocity based
on the clustering of cooperators \cite{Space1,Space2,Space3}. In the
latter case, the level of cooperation in two-dimensional spatial
games is further enhanced by ``disordered environments''
(approximately 10\% unaccessible empty locations) \cite{Vain1}, and
by diffusive mobility, provided that the mobility parameter is in a
suitable range \cite{Mobility1}. However, strategy mutations, random
relocations, and other sources of stochasticity (``noise'') can
significantly challenge the formation and survival of cooperative
clusters. When no mobility or undirected, random mobility are
considered, the level of cooperation in the spatial games studied by
us is sensitive to noise (see Figs. 1d and 3c), as favorable
correlations between cooperative neighbors are destroyed. {\it
Success-driven} migration, in contrast, is a robust mechanism: By
leaving unfavorable neighborhoods, seeking more favorable ones, and
remaining in cooperative neighborhoods, it supports cooperative
clusters very efficiently against the destructive effects of noise,
thus preventing defector invasion in a large area of payoff
parameters.
\par
We assume $N$ individuals on a square lattice with periodic boundary
conditions and $L\times L$ sites, which are either empty or occupied
by one individual. Individuals are updated asynchronously, in a
random sequential order. The randomly selected individual performs
simultaneous interactions with the $m=4$ direct neighbors and
compares the overall payoff with that of the $m$ neighbors.
Afterwards, the strategy of the best performing neighbor is copied
with probability $1-r$ (``imitation''), if the own payoff was lower.
With probability $r$, however, the strategy is randomly ``reset'':
\textbf{Noise 1} assumes that an individual spontaneously chooses to
cooperate with probability $q$ or to defect with probability $1-q$
until the next strategy change. The resulting strategy mutations may
be considered to reflect deficient imitation attempts or
trial-and-error behavior. As a side effect, such noise leads to an
independence of the finally resulting level of cooperation from the
initial one at $t=0$, and a {\it qualitatively different} pattern
formation dynamics for the same payoff values, update rules, and
initial conditions (see Fig. \ref{S1}). Using the alternative Fermi
update rule \cite{Space2} would have been possible as well. However,
resetting strategies rather than inverting them, combined with
values $q$ much smaller than 1/2, has here the advantage of creating
particularly adverse conditions for cooperation, independently of
what strategy prevails. Below, we want to learn, if predominant
cooperation can survive or even emerge under such adverse
conditions.
\par
``Success-driven migration'' has been implemented as follows
\cite{Hegselmann,ACS}: Before the imitation step, an individual
explores the expected payoffs for the empty sites in the migration
neighborhood of size $(2M+1)\times (2M+1)$ (the Moore neighborhood
of range $M$). If the fictitious payoff is higher than in the
current location, the individual is assumed to move to the site with
the highest payoff and, in case of several sites with the same
payoff, to the closest one (or one of them); otherwise it stays put.

\section{Results}
\par
Computer simulations of the above model show that, in the {\it
imitation-only} case of classical spatial games with noise 1, but
{\it without} a migration step, the resulting fraction of
cooperators in the PD tends to be very low. It basically reflects
the fraction $rq$ of cooperators due to strategy mutations. For
$r=q=0.05$, we find almost frozen configurations, in which only a
small number of cooperators survive (see Fig. \ref{S1}d). In the
{\it migration-only} case without an imitation step, the fraction of
cooperators changes only by strategy mutations. Even when the
initial strategy distribution is uniform, one observes the formation
of spatio-temporal patterns, but the patterns get almost frozen
after some time  (see Fig. \ref{S1}e).
\par
It is interesting that, although for the connectivity structure of
our PD model neither imitation only (Fig. \ref{S1}d) nor migration
only (Fig. \ref{S1}e) can promote cooperation under noisy
conditions, their {\it combination}  does: Computer simulations show
the formation of cooperative clusters with a few defectors at their
boundaries (see Fig. \ref{S1}f). Once cooperators are organized in
clusters, they tend to have more neighbors and to reach higher
payoffs on average, which allows them to survive
\cite{Epstein,Hegselmann,ACS}. It will now have to be revealed, how
success-driven migration causes the {\it formation} of clusters at
all, considering the opposing noise effects. In particular, we will
study, why defectors fail to invade cooperative clusters and to
erode them from within, although a cooperative environment is most
attractive to them.
\par
To address these questions, Figure 2 studies a ``defector's
paradise'' with a single defector in the center of a cooperative
cluster. In the noisy {\it imitation-only} spatial prisoner's
dilemma, defection tends to spread up to the boundaries of the
cluster, as cooperators imitate more successful defectors (see Figs.
2a-d). But if imitation is combined with {\it success-driven
migration}, the results are in sharp contrast: Although defectors
still spread initially, cooperative neighbors  who are $M$ steps
away from the boundary of the cluster can now evade them. Due to
this defector-triggered migration, the neighborhood reconfigures
itself adaptively. For example, a large cooperative cluster may
split up into several smaller ones (see Figs. 2e-h). Eventually, the
defectors end up at the boundaries of these cooperative clusters,
where they often turn into cooperators by imitation of more
successful cooperators in the cluster, who tend to have more
neighbors. This promotes the spreading of cooperation
\cite{Epstein,Hegselmann,ACS}.
\par
Since evasion takes time, cooperative clusters could still be
destroyed when continuously challenged by defectors, as it happens
under noisy conditions. Therefore, let us now study the effect of
different kinds of randomness \cite{Young,Epstein}. \textbf{Noise 1}
(defined above) assumes {\it strategy mutations}, but leaves the
spatial distribution of individuals unchanged (see Fig. 3a).
\textbf{Noise 2}, in contrast, assumes that individuals, who are
selected with probability $r$, move to a randomly chosen free site
without considering the expected success {\it (random relocations).}
Such random moves may potentially be of long distance and preserve
the number of cooperators, but have the potential of destroying
spatial patterns (see Fig. 3b). \textbf{Noise 3} combines noise 1
and noise 2, assuming that individuals randomly relocate with
probability $r$ and additionally reset their strategy as in noise 1
(see Fig. 3c).
\par
While cooperation in the imitation-only case is quite sensitive to
noise (see Figs. 3a-c), the combination of imitation with
success-driven motion is not (see Fig. 3d-f): Whenever an empty site
inside a cluster of cooperators occurs, it is more likely that the
free site is entered by a cooperator than by a defector, as long as
cooperators prevail within the migration range $M$. In fact, the
formation of small cooperative clusters was observed for {\it all}
kinds of noise. That is, the combination of imitation with
success-driven migration is a robust mechanism to maintain and even
spread cooperation under various conditions, given there are enough
cooperators in the beginning.
\par
It is interesting, whether this mechanism is also able to facilitate
a spontaneous {\it outbreak} of predominant cooperation in a noisy
world  dominated by selfishness, without a ``shadow of the future''
\cite{Axelrod,Huberman}. Our simulation scenario assumes defectors
only in the beginning (see Fig. 4a), strategy mutations in favor of
defection, and short-term payoff-maximizing behavior in the vast
majority of cases. In order to study conditions under which a
significant fraction of cooperators is unlikely, our simulations are
performed with noise 3 and $r=q=0.05$, as it tends to destroy
spatial clusters and cooperation (see Fig. 3c):  By relocating 5\%
randomly chosen individuals in each time step, noise 3 dissolves
clusters into more or less separate individuals in the
imitation-only case (see Figs. 3b+c). In the case with
success-driven migration, random relocations break up large clusters
into many smaller ones, which are distributed all over the space
(see Figs. 3b+c and 4b). Therefore, even the clustering tendency by
success-driven migration can only partially compensate for the
dispersal tendency by random relocations. Furthermore, the strategy
mutations involved in noise 3 tend to destroy cooperation (see Figs.
3a+c, where the strategies of 5\% randomly chosen individuals were
replaced by defection in 95\% of the cases and by cooperation
otherwise, to create conditions favoring defection, i.e. the
dominant strategy in the prisoner's dilemma). Overall, as a result
of strategy mutations (i.e. without the consideration of imitation
processes), only a fraction $rq = 0.0025$ of all defectors turn into
cooperators in each time step, while a fraction $r(1-q)\approx 0.05$
of all cooperators turn into defectors (i.e. 5\% in each time step).
This setting is extremely unfavorable for the spreading of
cooperators. In fact, defection prevails for an extremely long time
(see Figs. 4b and \ref{S2}a). But suddenly, when a small,
supercritical cluster of cooperators has occurred by coincidence
(see Fig. 4c), the fraction of cooperators spreads quickly (see Fig.
\ref{S2}a), and soon cooperators prevail (see Figs. 4d and
\ref{S2}b). Note that this spontaneous birth of predominant
cooperation in a world of defectors does not occur in the noisy
imitation-only case and demonstrates that success-driven migration
can overcome the dispersive tendency of noises 2 and 3, if $r$ is
moderate and $q$ has a finite value. That is, success-driven
migration generates spatial correlations between cooperators more
quickly than these noises can destroy them. This changes the outcome
of spatial games essentially, as a comparison of Figs. 2a-d with
4a-d shows.
\par
The conditions for the spreading of cooperators from a supercritical
cluster (``nucleus'') can be understood by configurational analysis
\cite{Clusters,Young} (see Fig. {\bf S1}), but the underlying
argument can be both, simplified and extended: According to Fig. 6a,
the level of cooperation changes when certain lines (or, more
generally, certain hyperplanes) in the payoff-parameter space are
crossed. These hyperplanes are all of the linear form
\begin{equation}
n_1 R + n_2 S = n_3 T + n_4 P \, , \label{eins}
\end{equation}
where $n_k \in \{0,1,2,3,4\}$. The left-hand side of Eq. [1]
represents the payoff of the most successful cooperative neighbor of
a focal individual, assuming that this has $n_1$ cooperating and
$n_2$ defecting neighbors, which implies $n_1+n_2 \le m=4$. The
right-hand side reflects the payoff of the most successful defecting
neighbor, assuming that $n_3$ is the number of his/her cooperating
neighbors and $n_4$ the number of defecting neighbors, which implies
$n_3+n_4 \le m=4$. Under these conditions, the best-performing
cooperative neighbor earns a payoff of $n_1 R + n_2 S$, and the
best-performing defecting neighbor earns a payoff of $n_3 T + n_4
P$. Therefore, the focal individual will imitate the cooperator, if
$n_1 R + n_2 S > n_3 T + n_4 P$, but copy the strategy of the
defector if $n_1 R + n_2 S < n_3 T + n_4 P$. Equation [1] is the
line separating the area where cooperators spread (above the line)
from the area of defector invasion (below it) for a certain spatial
configuration of cooperators and defectors  (see Fig. 6a). Every
spatial configuration is characterized by a set of $n_k$-parameters.
As expected, the relative occurence frequency of each configuration
depends on the migration range $M$ (see Fig. 6b): Higher values of
$M$ naturally create better conditions for the spreading of
cooperation, as there is a larger choice of potentially more
favorable neighborhoods.
\par
Figure 6b also shows that success-driven migration extends the
parameter range, in which cooperators prevail, from the parameter
range of the snowdrift game with $S>P$ to a considerable parameter
range of the prisoner's dilemma. For this to happen, it is important
that the attraction of cooperators is mutual, while the attraction
of defectors to cooperators is not. More specifically, the
attraction of cooperators is proportional to $2R$, while the
attraction between defectors and cooperators is proportional to
$T+S$. The attraction between cooperators is stronger, because the
prisoner's dilemma usually assumes the inequality $2R > T+S$.
\par
Besides the speed of finding neighbors to interact with, the time
scales of configurational changes and correlations matter as well:
By entering a cooperative cluster, a defector triggers an avalanche
of strategy changes and relocations, which quickly destroys the
cooperative neighborhood. During this process, individuals may alter
their strategy many times, as they realize opportunities by
cooperation or defection immediately. In contrast, if a cooperator
joins a cooperative cluster, this will stabilize the cooperative
neighborhood. Although cooperative clusters continuously adjust
their size and shape,  the average time period of their existence is
longer than the average time period after which individuals change
their strategy or location.  This coevolution of social interactions
and strategic behavior reflects features of many social
environments: While the latter come about by individual actions, a
suitable social context can make the average behavior of individuals
more predictable, which establishes a reinforcement process. For
example, due to the clustering tendency of cooperators, the
likelihood of finding another cooperator in the neighborhood of a
cooperator is greater than 1/2, and also the likelihood that a
cooperator will cooperate in the next iteration.

\section{Discussion}
\par
\begin{figure*}[t]
\begin{center}
    \includegraphics[width=8.7cm]{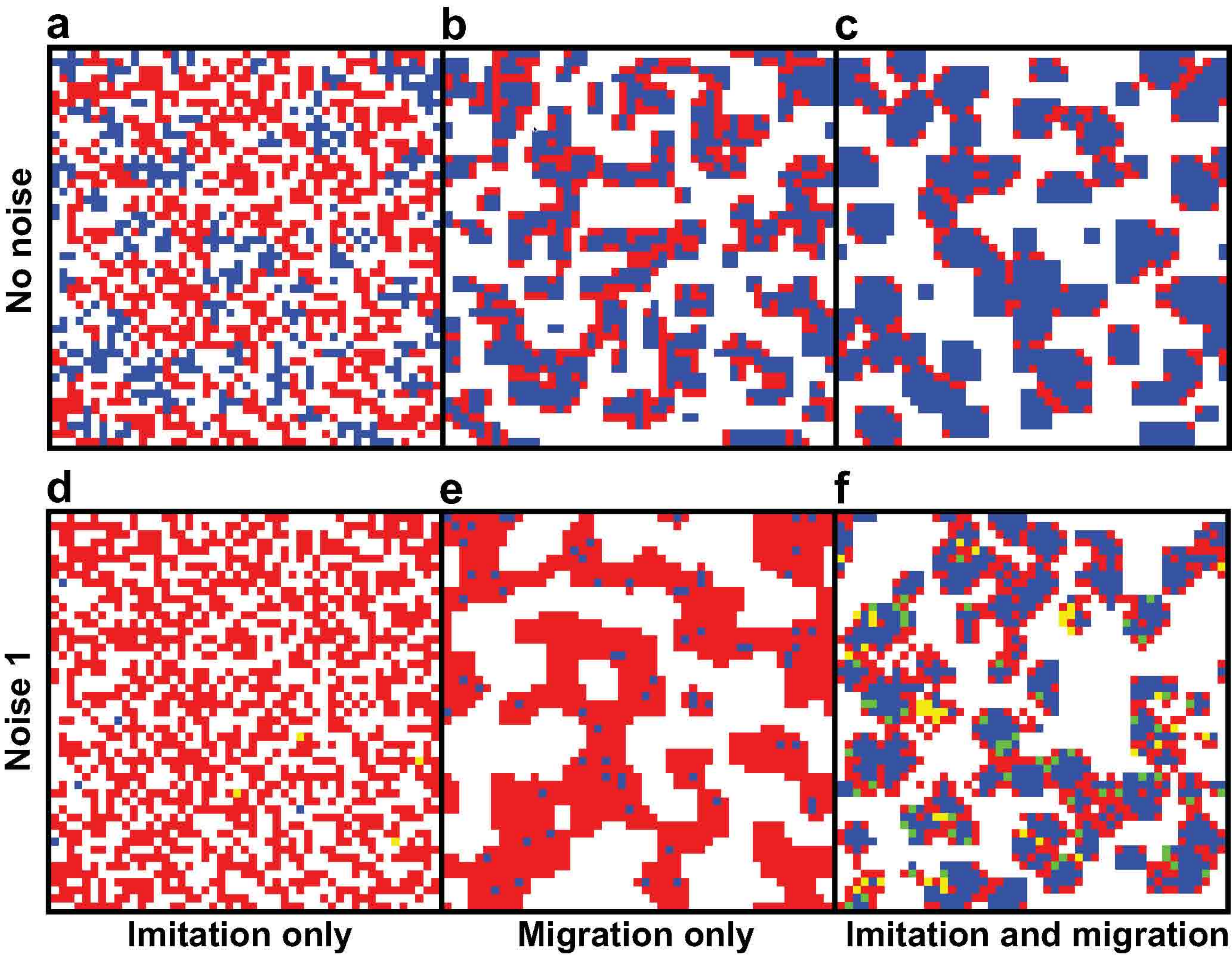}
\end{center}
\caption{\sffamily Representative simulation results  for the
spatial prisoner's dilemma with payoffs $T=1.3$, $R=1$, $P=0.1$, and
$S=0$ after $t=200$ iterations. The simulations are for  $49\times
49$-grids with 50\% empty sites. At time $t=0$ we assumed 50\% of
the individuals to be cooperators and 50\% defectors. Both
strategies were homogeneously distributed over the whole grid. For
reasons of comparison, all simulations were performed with identical
initial conditions and random numbers (red = defector, blue =
cooperator, white = empty site, green = defector who became a
cooperator in the last iteration, yellow = cooperator who turned
into a defector). Compared to simulations without noise (top), the
strategy mutations of noise 1 with $r=q=0.05$ do not only reduce the
resulting level of cooperation, but also the outcome and pattern
formation dynamics, even if the payoff values, initial conditions,
and update rules are the same (bottom): In the imitation-only case
with $M=0$ that is displayed on the left, the initial fraction of
50\% cooperators is quickly reduced due to imitation of more
successful defectors. The result is a ``frozen'' configuration
without any further strategy changes. (a) In the noiseless case, a
certain number of cooperators can survive in small cooperative
clusters. (d) When noise 1 is present, random strategy mutations
destroy the level of cooperation almost completely, and the
resulting level of defection reaches values close to 100\%. The
illustrations in the center show the migration-only case with
mobility range $M=5$: (b) When no noise is considered, small
cooperative clusters are formed, and defectors are primarily located
at their boundaries. (e) In the presence of noise 1, large clusters
of defectors are formed instead, given $P>0$. The illustrations on
the right show the case, where imitation is combined with
success-driven migration (here, $M=5$): (d) In the noiseless case,
cooperative clusters grow and eventually freeze (i.e. strategy
changes or relocations do not occur any longer). (f) Under noisy
conditions, in contrast, the cooperative clusters continue to adapt
and reconfigure themselves, as the existence of yellow and green
sites indicates.} \label{S1}
\end{figure*}

It is noteworthy that all the above features---the survival of
cooperation in a large parameter area of the PD,  spatio-temporal
pattern formation,  noise-resistance, and the outbreak of
predominant cooperation---can be captured by considering a mechanism
as simple as success-driven migration: Success-driven migration {\it
destabilizes} a homogeneous strategy distribution (compare Fig.
\ref{S1}c with \ref{S1}a and Fig. \ref{S1}f with \ref{S1}d). This
triggers the spontaneous formation of agglomeration and segregation
patterns \cite{Schelling}, where noise or diffusion would cause
dispersal in the imitation-only case. The self-organized patterns
create self-reinforcing social environments characterized by
behavioral correlations, and imitation promotes the further growth
of supercritical cooperation clusters. While each mechanism by
itself tends to produce frozen spatial structures, the combination
of imitation and migration supports adaptive patterns (see Fig.
\ref{S1}f). This facilitates, for example, the regrouping of a
cluster of cooperators upon invasion by a defector, which is crucial
for the survival and success of cooperators (see Fig. 2e-h).
\par
By further simulations we have checked that our conclusions are
robust with respect to using different update rules, adding birth
and death processes, or introducing a small fraction of individuals
defecting unconditionally. The same applies to various kinds of
``noise''. Noise can even trigger cooperation in a world full of
defectors, when the probability of defectors to turn spontaneously
into cooperators is 20 times smaller than the probability of
cooperators to turn into defectors. Compared to the implications of
the game-dynamical replicator equation, this is remarkable: While
the replicator equation predicts that the stationary solution with a
majority of cooperators is {\it un}stable with respect to
perturbations and the stationary solution with a majority of
defectors is stable \cite{Epstein}, success-driven migration {\it
inverts} the situation: The state of 100\% defectors becomes {\it
unstable} to noise, while a majority of cooperators is stabilized in
a considerable area of the payoff parameter space.
\par
Our results help to explain why cooperation can be frequent even if
individuals would behave selfishly in the vast majority of
interactions. Although one may think that migration would weaken
social ties and cooperation, there is another side of it which helps
to establish cooperation in the first place, without the need to
modify the payoff structure. We suggest that, besides the ability
for strategic interactions and learning, the ability to {\it move}
has played a  crucial role for the evolution of large-scale
cooperation and social behavior. Success-driven migration can reduce
unbalanced social interactions, where cooperation is unilateral, and
support local agglomeration. In fact, it has been pointed out that
local agglomeration is an important precondition for the evolution
of more sophisticated kinds of cooperation \cite{Deneubourg}. For
example, the level of cooperation could be further improved by
combining imitation and success-driven migration with other
mechanisms such as costly punishment \cite{Punish,Gintis1},
volunteering \cite{Space2}, or reputation
\cite{Scoring,Partner,Rockenbach}.

\begin{acknowledgments}
{\small The authors would like to thank Christoph Hauert, Heiko
Rauhut, Sergi Lozano, Michael Maes, Carlos P. Roca, and Didier
Sornette for their comments.}
\end{acknowledgments}



\end{article}

%
%
%
%

\newpage

\begin{figure*}[t]
\begin{center}
\includegraphics[width=8.7cm]{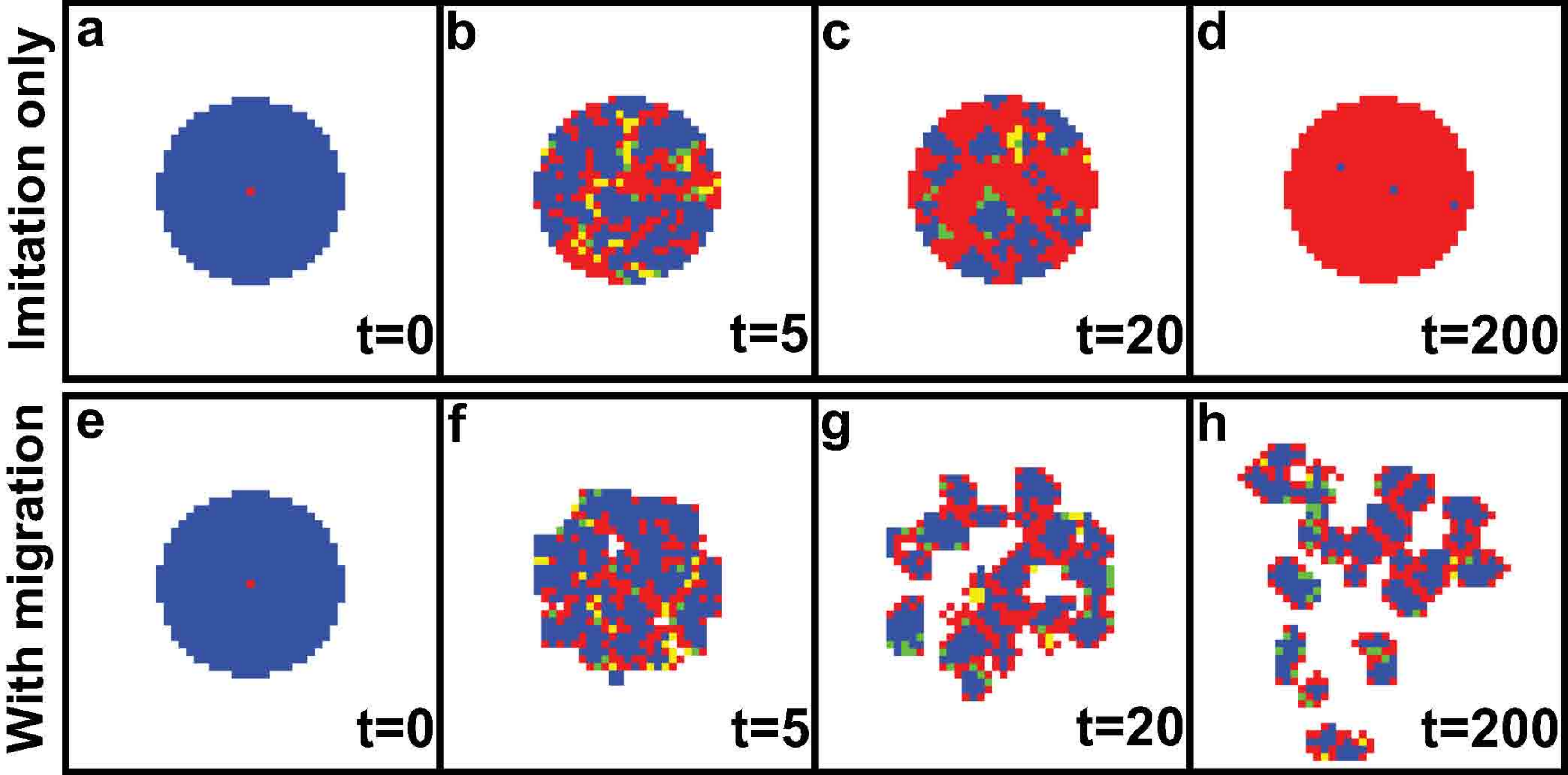}
\end{center}
    \caption{\sffamily Representative simulation results  after $t=200$ iterations in the ``defector's paradise'' scenario, starting with a single defector in the center of a cooperative cluster at $t=0$. The simulations are performed on $49\times 49$-grids with $N=481$ individuals, corresponding to a circle of diameter 25. They are based on the spatial prisoner's dilemma with payoffs $T=1.3$, $R=1$, $P=0.1$, $S=0$ and noise parameters $r=q=0.05$ (red = defector, blue = cooperator, white = empty site, green = defector who became a cooperator, yellow = cooperator who turned into a defector in the last iteration). For reasons of comparison, all simulations were carried out with identical initial conditions and random numbers.
(a-d) In the noisy imitation-only case with $M=0$, defection (red)
eventually spreads all over the cluster. The few remaining
cooperators (blue) are due to strategy mutations. (e-h) When we add
success-driven motion, the result is very different. Migration
allows cooperators to evade defectors. That triggers a splitting of
the cluster, and defectors end up on the surface of the resulting
smaller clusters, where most of them can be turned into cooperators.
This mechanism is crucial for the unexpected survival and spreading
of cooperators. } \label{figure1}
\end{figure*}

\begin{figure*}[t]
\begin{center}
    \includegraphics[width=8.7cm]{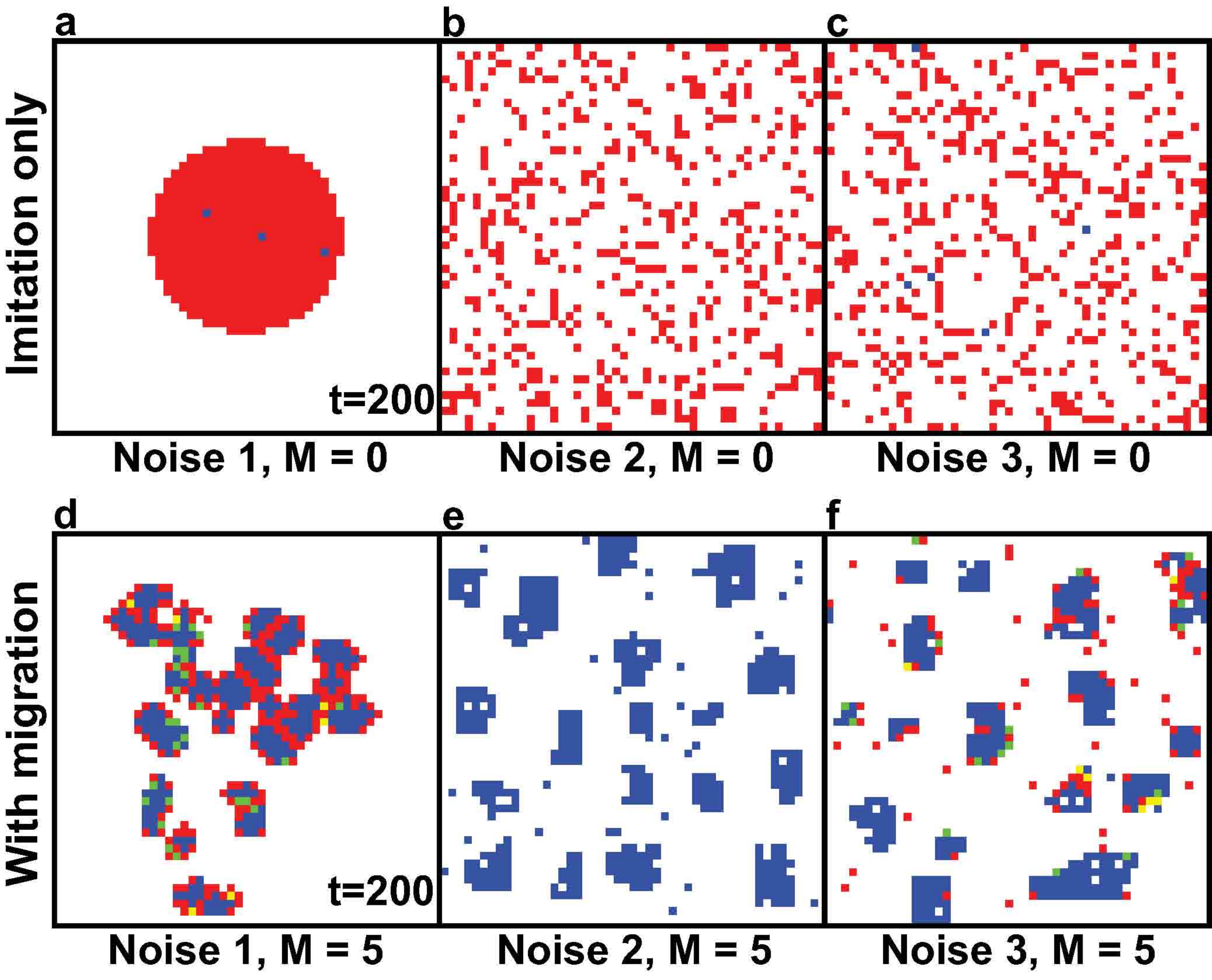}
\end{center}
\caption{\sffamily Representative simulation results for the
invasion scenario with a defector in the center of a cooperative
cluster (``defector's paradise''). The chosen payoffs $T=1.3$,
$R=1$, $P=0.1$, and $S=0$ correspond to a prisoner's dilemma. The
simulations are for  $49\times 49$-grids with $N=481$ individuals,
corresponding to a circle of diameter 25 (red = defector, blue =
cooperator, white = empty site, green = defector who became a
cooperator, yellow = cooperator who turned into a defector in the
last iteration). Top: Typical numerical results for the
imitation-only case ($M=0$) after $t=200$ iterations (a) for noise 1
(strategy mutations) with mutation rate $r=0.05$ and creation of
cooperators with probability $q=0.05$, (b) for noise 2 (random
relocations) with relocation rate $r=0.05$, and (c) for noise 3 (a
combination of random relocations and strategy mutations) with
$r=q=0.05$. As cooperators imitate defectors with a higher overall
payoff, defection spreads easily. The different kinds of noise
influence the dynamics and resulting patterns considerably: While
strategy mutations in (a) and (c) strongly reduce the level of
cooperation, random relocations in (b) and (c) break up spatial
clusters, leading to a dispersion of individuals in space. Their
combination in case (c) essentially destroys both, clusters and
cooperation. Bottom: Same for the case of imitation \textit{and}
success-driven migration with mobility range $M=5$ (d) for noise 1
with $r=q=0.05$, (e) for noise 2 with $r=0.05$, and (f) for noise 3
with $r=q=0.05$. Note that noise 1 just mutates strategies and does
not support a spatial spreading, while noise 2 causes random
relocations, but does not mutate strategies. This explains why the
clusters in Fig. 3d do not spread out over the whole space and why
no new defectors are created in Fig. 3e. However, the creation of
small cooperative clusters is found in all three scenarios.
Therefore, it is robust with respect to various kinds of noise, in
contrast to the imitation-only case.} \label{figure2}
\end{figure*}

\begin{figure*}[t]
\begin{center}
    \includegraphics[width=14cm]{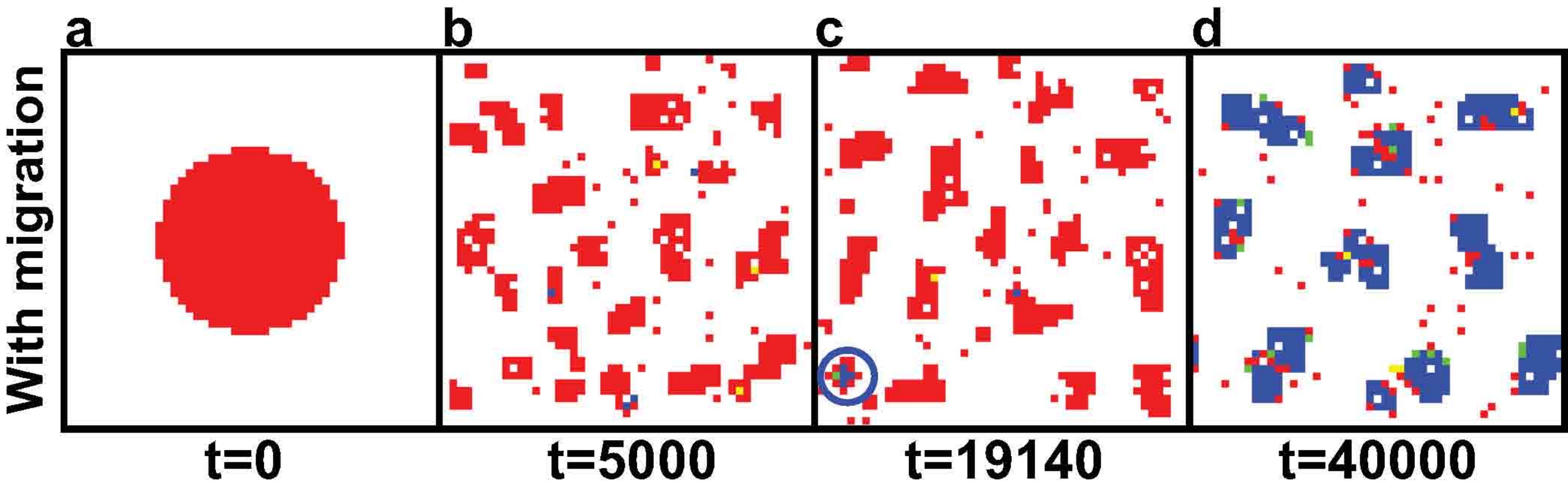}
\end{center}
\caption{\sffamily Spontaneous outbreak of prevalent cooperation in
the spatial prisoner's dilemma with payoffs $T=1.3$, $R=1$, $P=0.1$,
$S=0$ in the presence of noise 3 (random relocations and strategy
mutations) with $r=q=0.05$. The simulations are for  $49\times
49$-grids  (red = defector, blue = cooperator, white = empty site,
green = defector who became a cooperator, yellow = cooperator who
turned into a defector in the last iteration). (a) Initial cluster
of defectors, which corresponds to the \textit{final} stage of the
\textit{imitation-only} case with strategy mutations according to
noise 1 (see Fig. 2d). (b) Dispersal of defectors by noise 3, which
involves random relocations. A few cooperators are created randomly
by strategy mutations with the very small probability $rq=0.0025$
(0.25 percent). (c) Occurrence of a supercritical cluster of
cooperators after a very long time. This cooperative ``nucleus''
originates by random coincidence of favorable strategy mutations in
neighboring sites. (d) Spreading of cooperative clusters in the
whole system. This spreading despite the destructive effects of
noise requires an effective mechanism to form growing cooperative
clusters (such as success-driven migration) and cannot be explained
by random coincidence. See the supplementary video for an animation
of the outbreak of cooperation for a different initial condition.}
\label{figure3}
\end{figure*}

\begin{figure*}[t]
\begin{center}
\includegraphics[width=7cm]{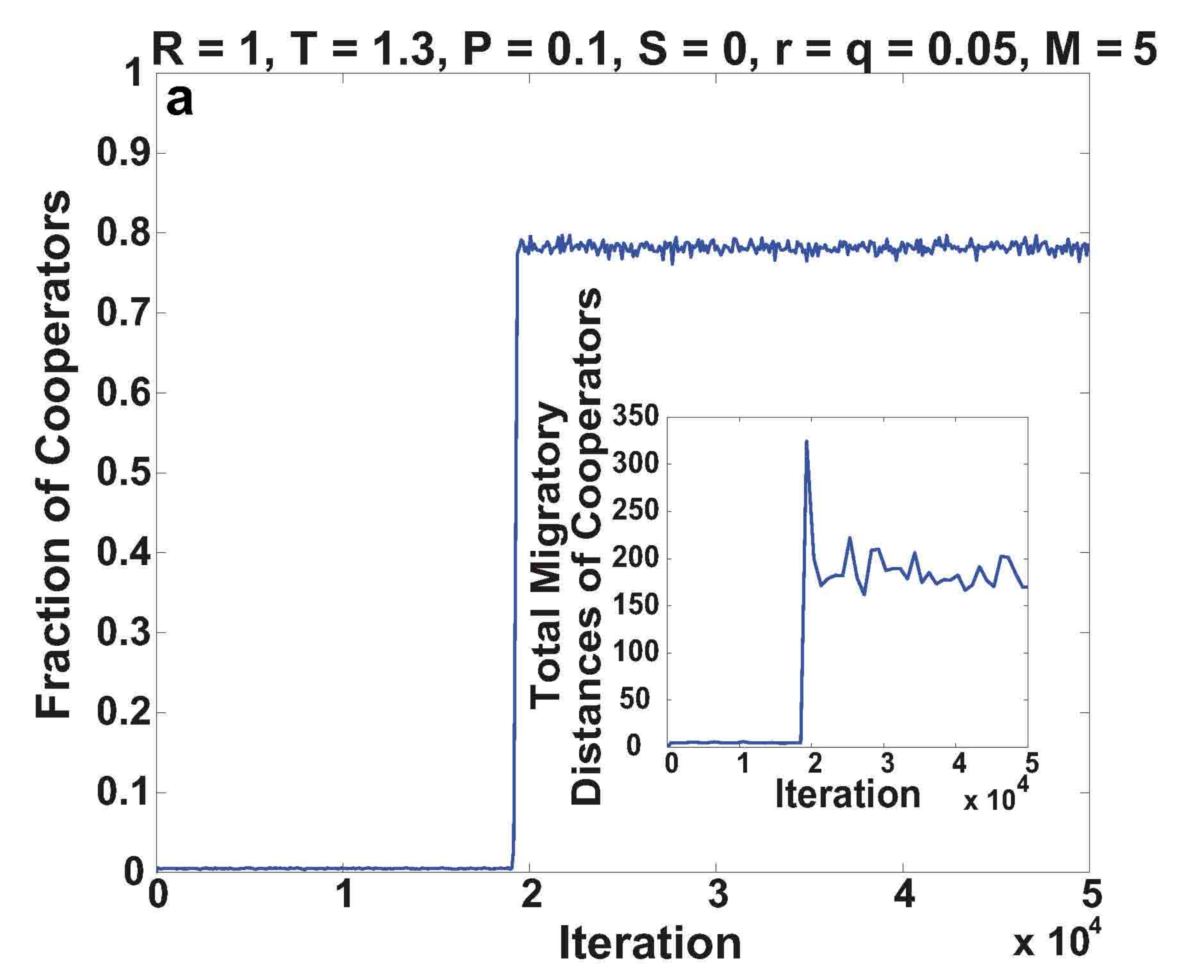}
\includegraphics[width=6.8cm]{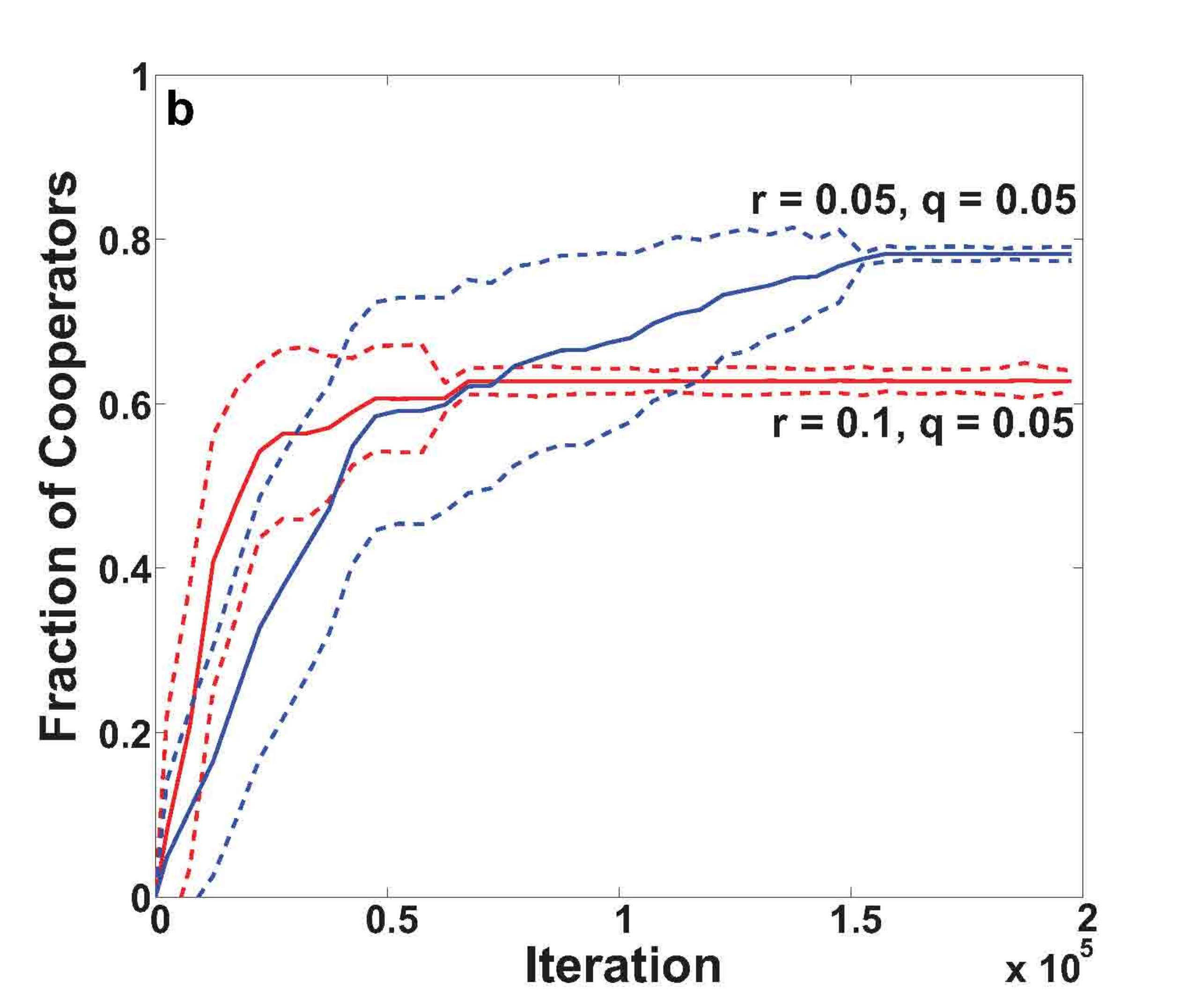}
\end{center}
\caption{\sffamily Representative example for the outbreak of
predominant cooperation in the prisoner's dilemma with payoffs
$T=1.3$, $R=1$, $P=0.1$, $S=0$, in the presence of noise 3 with
$r=q=0.05$. The simulations are for  $49\times 49$-grids  with a
circular cluster of defectors and no cooperators in the beginning
(see Fig. 4a). (a) After defection prevails for a very long time
(here for almost 20,000 iterations), a sudden transition to a large
majority of cooperators is observed. Inset: The overall distance
moved by all individuals during one iteration has a peak at the time
when the outbreak of cooperation is observed. Before, the rate of
success-driven migration is very low, while it stabilizes at an
intermediate level afterwards. This reflects a continuous evasion of
cooperators from defectors and, at the same time, the  continuous
effort to form and maintain cooperative clusters. The graph displays
the amount of success-driven migration only, while the effect of
random relocations is not shown. (b) Evaluating 50 simulation runs,
the error bars (representing 3 standard deviations) show a large
variation of the time points when prevalent cooperation breaks out.
Since this time point depends on the coincidence of random
cooperation in neighboring sites, the large error bars have their
natural reason in the stochasticity of this process. After a
potentially very long time period, however, all systems end up with
a high level of cooperation. The level of cooperation decreases with
the noise strength $r$, as expected, but moderate values of $r$ can
even \textit{accelerate} the transition to predominant cooperation.
Using the parameter values $r=0.1$ and $q=0.2$, the outbreak of
prevalent cooperation takes often less than 200 iterations.}
\label{S2}
\end{figure*}


\begin{figure*}[t]
\begin{center}
    \includegraphics[width=7cm]{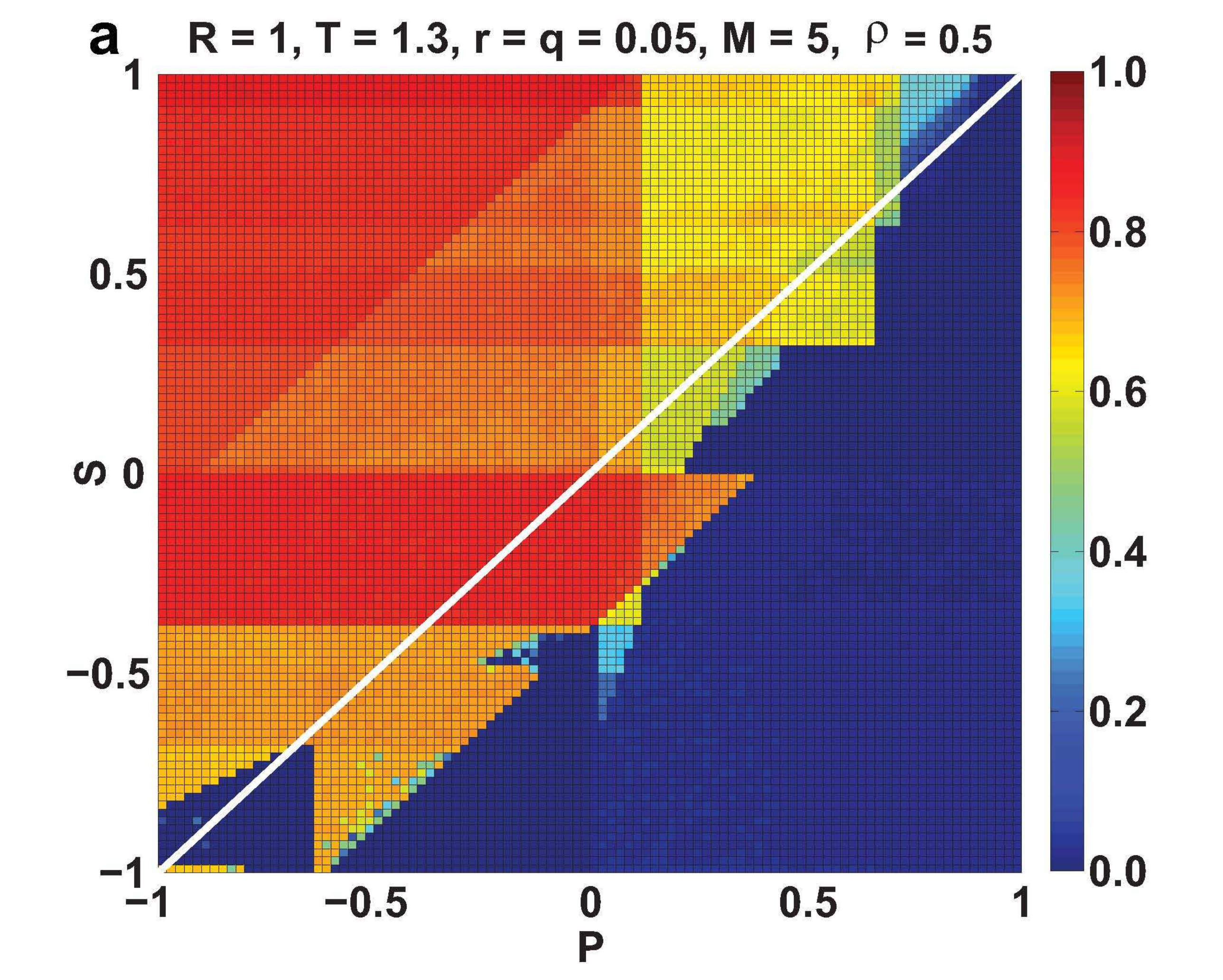}
    \includegraphics[width=6.75cm]{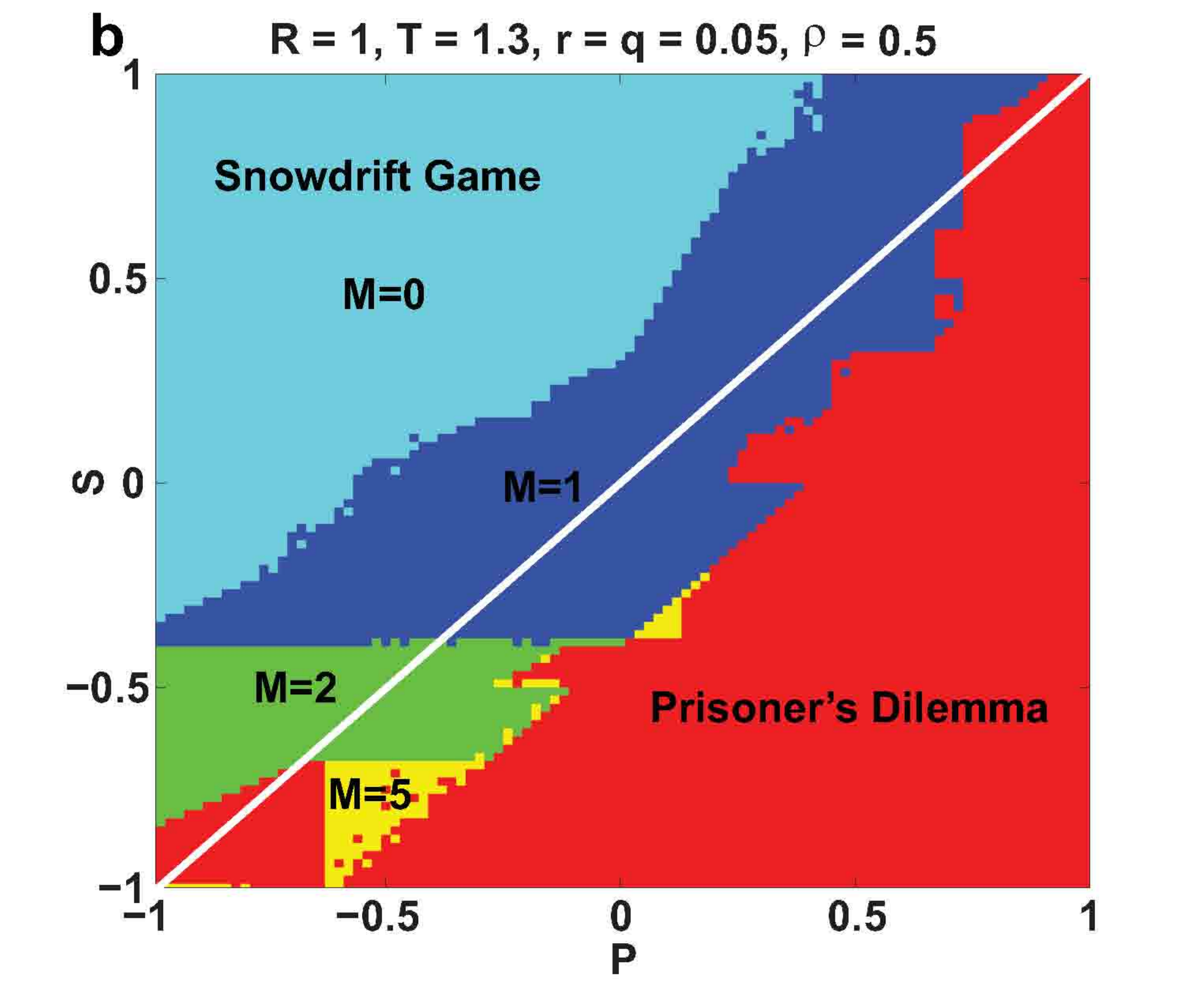}
\end{center}
\caption{\sffamily Dependence of the fraction of cooperators for
given payoff parameters $T=1.3$ and $R=1$ on the parameters $P$ and
$S$. The area above the solid diagonal line corresponds to the
snowdrift game, the area below to the prisoner's dilemma. Our
simulations were performed for grids with $L\times L = 99\times 99$
sites and $N=L^2/2$ individuals, corresponding to a density
$\rho=N/L^2 = 0.5$. At time $t=0$ we assumed 50\% of the individuals
to be cooperators and 50\% defectors. Both strategies were
homogeneously distributed over the whole grid. The finally resulting
fraction of cooperators was averaged at time $t=200$ over 50
simulation runs  with different random realizations. The simulations
were performed with noise 3 (random relocations with strategy
mutations) and $r=p=0.05$. An enhancement in the level of
cooperation (often by more than 100\%) is observed mainly in the
area with $P-0.4 < S < P + 0.4$ and $P < 0.7$. Results for the
noiseless case with $r=0$ are shown in Fig. \textbf{S2}. (a) The
fraction of cooperators is represented by color codes (see the bar
to the right of the figure, where dark orange, for example,
corresponds to 80\% cooperators). It can be seen that the fraction
of cooperators is approximately constant in areas limited by
straight lines (mostly triangular and rectangular ones). These lines
correspond to Eq. [1] for different specifications of $n_1$, $n_2$,
$n_3$, and $n_4$ (see main text for details). (b) The light blue
area reflects the parameters for which cooperators reach a majority
in the imitation-only case with $M=0$: For all payoffs $P$ and $S$
corresponding to a prisoner's dilemma, cooperators are clearly in
the minority, as expected. However, taking into account
success-driven migration changes the situation in a pronounced way:
For a mobility range $M=1$, the additional area with more than 50\%
cooperators is represented by dark blue, the further extended area
of prevailing cooperation for $M=2$ by green color, and for $M=5$ in
yellow. If $M=5$, defectors are in the majority only for parameter
combinations falling into the red area. This demonstrates that
success-driven migration can promote predominant cooperation in
considerable areas, where defection would prevail without migration.
For larger interaction neighborhoods $m$, e.g. $m=8$, the area of
prevalent cooperation is further increased overall (not shown). Note
that the irregular shape of the separating lines is no artefact of
the computer simulation or initial conditions. It results by
superposition of the areas defined by Eq. [1], see Fig. 6a. }
\label{figure4}
\end{figure*}

\end{document}